\documentclass[twocolumn,showpacs,superscriptaddress,pre]{revtex4}
\usepackage{graphicx}
\usepackage{amsmath,amssymb}
\usepackage{hyperref}
\begin{document}
\title{Deriving The Bekenstein-Hawking entropy by Statistical Mechanics and Quantization of Non-Commuting Space}
\author{R. Hassannejad}
\email{r.hassannejad@ut.ac.ir}
\affiliation{Department of Physics, University of Tehran, Tehran 14395-547, Iran}

\date{\today}

\begin{abstract}
In this paper we are going to derive the Bekenstein-Hawking entropy by considering a Non-Commuting two dimensional quantized space, and we will show that the Bekenstein-Hawking entropy is valid for a system (black hole) in an equilibrium state. In addition, by using the entropy was obtained, we derive the Einstein and Newton gravity equations and finally we get an equation that is consistent with the previous results. In other words, we want to show that by using the statistical mechanics and quantization of space and the excitation of space quantas, we will be able to obtaine the Einstein and Newton gravity equations. In general, we try to show that all equations of gravity and the thermodynamics of gravitational systems have quantum and statistical origin.
\end{abstract}
\maketitle

Studying the black body radiation thermodynamics by Planck, Einstein, Boltzmann and  others  led to the formulation of quantum mechanics. and beside that, Einstein formulated the special and general relativity. Although, both of these theories have had many achievements but when we want to combine the two theories to have a unique theory to describing the world, we get a series of mathematical inconsistencies. Historical experiences tell maybe thermodynamics  helps us to solve this problem. The relation between gravity and thermodynamics of gravitational systems was introduced by Hawking, Bekenstein and others between 1972 and 1985 in series of papers \cite{1,2,3,4,5,6,7,8,9,10,11,12,13,14,15,16,17}. In another case, Jacobson tried to derive Einstein's equations of gravity from the proportionality of entropy and horizon area and by using the  Clausius equation \cite{18} or in 2011 Verlinde derived the Einstein's and Newton's equations of gravity from the concept of entropic force \cite{19}. Besides, in the past few decades many people have tried to solve  that mathematical inconsistencies by other ways, and many theories have been developed to solve this problem. Two of the  most successful of these theories are string theory and loop quantum gravity. Also, the holographic principle was presented by Hooft \cite{20} and Susskind \cite{21} paved the way for a series of advanced studies on gravity and other fundamental forces, for example by the Maldacena work in 1998,\cite{22}, theoretical physics has experienced a major change. In addition to the works have been done to understanding the nature of gravity, many people also tried to understand the origin of the black hole entropy. For example, In 1996 Rovelli tried to obtain the Bekenstein-Hawking entropy by using loop quantum gravity. and finaly, The equation that was obtained by him about the entropy of black holes was only different in one coefficient than Bekenstein-Hawking entropy \cite{23} (In addition to Rovelli 's paper, several papers have been written about the relationship between quantum gravity and black hole entropy) \cite{24,25,26}. Along with Rovelli, in 1996 Vafa and Strominger attempted to obtain the Bekenstein-Hawking entropy by using a model of string theory \cite{27}. Apart from the work has been done to understanding the nature of the  black hole entropy  in this paper, we want to know how to obtain black hole entropy by quantization of space. To do this, just by  following the procedure provided in reference\cite{28}. We can get the area eigenvalues of space quantas, by the help of the operators of creation and annihilation, and assuming that our two-dimensional space is Non-Commuting 
\begin{equation}\label{eq1}
A_n = 2 \hbar \Theta(n + \frac{1}{2})
\end{equation}
In this equation $n$ is assumed to be the number of space quantums. In paper \cite{29} the authors have assumed that n has the concept of state number rather than the number of space quantums. (Similar to a simple harmonic oscillator where n represents the energy states of the system, and the system is set to $n = 0$ at the ground state and by receiving energy it climbs to higher modes). And with the help of Eq .\eqref {eq1} they tried to find the equations governing gravity. The main idea of the authors in paper \cite{29} was to assume by inserting some energy or mass into the space its quantas would become excited and since the quantas of space is excited by energy on space so there must be a direct relationship between the energy absorbed by space quantas and Eq .\eqref {eq1} . Which in the paper \cite{29}, the authors have written this Relationship as $(E_n={\alpha}A_n,{\alpha}={\kappa}{\Omega})$ ($\kappa$ is an unknown dimensionless constant and $\Omega$ is a constant to set the dimension). In the following, we will say a brief description of the article \cite{29} and then we will use these calculations to obtain the Bekenstein-Hawking entropy. The average area of space quantas is obtained by the following equation
\begin{equation}\label{eq2}
\langle A \rangle = \frac{\sum_{n=0}^{\infty}A_n e^{-\beta A_n}}{\sum_{n=0}^{\infty}e^{-\beta A_n}} 
\end{equation}
Since in Eq .\eqref {eq2} the dimension of $A_n$  is  area and the argument ${(\beta A_n)}$  is dimensionless, so the dimension of $\beta$ must be the inverse of  area. For this reason we define $\beta = \frac{\alpha}{k_BT}$, where $T$ and $k_B$ are respectively the system temperature and the Boltzmann constant. It should be noted that in Eq .\eqref {eq2} the quantity $\sum_{n=0}^{\infty}e^{-\beta A_n}$  is the partition function of the system. With some simplification we  can written the  Eq .\eqref {eq2} as
\begin{equation}\label{eq3}
\langle A \rangle = \hbar \Theta \Big( 1 + \frac{2}{e^{2\beta(\hbar \Theta)}-1} \Big)
\end{equation}
We assume there are no interaction between space quantas. and so we can write the following relation between the total area of $N$ quantas and the average area of each quanta
\begin{equation}\label{eq4}
A_0 = N\langle A \rangle
\end{equation}
As stated, it was assumed that the each quanta of space did not interact with each other and so the partition function of the whole system is calculated as 
\begin{equation}\label{eq5}
Z = (Z_1)^N = \Big( \frac{e^{-\beta \hbar \Theta}}{1 - e^{-2\beta \hbar \Theta}} \Big)^N
\end{equation}
By using the partition function the entropy of system can be written as
\begin{equation}\label{eq6}
S = (\frac{k_B c^3 A_0}{\hbar G})\gamma(T)
\end{equation}
Where $\gamma(T)$  is defined as
\begin{equation}\label{eq7}
\gamma(T) = (\frac{b}{T}-{\Delta}),({b}={\kappa}{T_p})
\end{equation}
\begin{equation}\label{eq8}
\Delta = ( \frac{e^\frac{b}{T} -1}{e^\frac{b}{T} + 1}\Big) ln(e^\frac{b}{T} - e^\frac{-b}{T} )
\end{equation}
Where $T_p$  is Planck's temperature. It can be shown that in Eq \eqref {eq6} if we set the temperature to zero $(T \rightarrow0)$ at the same time entropy tends to zero$(S \rightarrow0)$. Besides, the entropy Eq \eqref {eq6} is a function of temperature and area in contrast to the Bekenstein-Hawking entropy of black holes, which is just a function of the area. But is there any connection between these two entropies? It may be possible to obtain the Bekenstein-Hawking entropy by using the Eq \eqref {eq6}, which we will discuss in the following. In the beginning, we will maximize Eq \eqref {eq6} for temperature (Assuming $A_0$ are constant or independent to the temperature
\begin{equation}\label{eq9}
\frac{\partial S}{\partial T}=0 \Rightarrow \ T = \frac{\kappa T_p}{ln(\frac{\sqrt{5}+1}{2})}
\end{equation}                          
If we put the temperature expression in Eq \eqref {eq9} to  Eq \eqref {eq6} the maximum entropy will be obtained, which is only a function of  area
\begin{equation}\label{eq10}
S_{max} = (\frac{k_B c^3 A_0}{\hbar G}){ln(\frac{\sqrt{5}+1}{2})}
\end{equation}
The entropy of the  Eq \eqref {eq10} is very similar to the Bekenstein-Hawking entropy, the only difference between the two equations are the coefficient of  Eq \eqref {eq10} $ln(\frac{\sqrt{5}+1}{2})$  rather than the coefficient  $(\frac{1}{4})$ in the Bekenstein-Hawking entropy. Where, the value of the maximum entropy coefficient is very close to the Bekenstein-Hawking entropy coefficient
\begin{equation}\label{eq11}
ln(\frac{\sqrt{5}+1}{2})   \simeq \ \frac{1}{2}  \Rightarrow \ S_{max}  \simeq \  (\frac{k_B c^3 A_0}{2\hbar G}),
\end{equation}
And in other words, we have been able to obtain the Bekenstein-Hawking entropy through the quantization of space and the idea of space quanta excitation. Now the question that comes to mind is: what is the origin of this entropy? And, why black holes have entropy? If the entropy of  Eq \eqref {eq10} is the entropy of the black hole event horizon, then it can be said that the origin of the entropy is the excitation of space quantas. where shown in Eq \eqref {eq1}. In other words, when some energy is in space, this energy is distributed between the quantas of space, but since space quantas are excited so energy can be divided in different ways between space quantas. And if we were able to count the number of these methods we would calculate the system entropy. so the source of the black hole entropy is our lack of awareness of the number of states that space quantas can get in the presence of some energy. For example, suppose some energy enters to the black hole's horizon, so each quantas of the event horizon will be excited but we do not know, what extent they are aroused? (Each of the space quantities can go up from their basic state $(n = 0)$ to the excited state distributed $n = 1,2,3, ...$). So with these descriptions, we should have entropy for the black hole event horizon and more generally for space-time. In the above equations, we showed that if we maximize Eq \eqref {eq6} for temperature, we read the Eq \eqref {eq10} which is very similar to the Bekenstein-Hawking entropy. Where, we know from thermodynamics and statistical mechanics a system is in equilibrium state when it is at its maximum entropy, so the entropy of Eq \eqref {eq11} predicts a system that is in thermodynamic equilibrium. In other words, the Bekenstein-Hawking entropy is valid only for a system that is in thermodynamic equilibrium. Further, Jacobson in 1995 showed that Einstein's gravity equation can be obtained from thermodynamic equilibrium with the help of Clausius equation, the full details of Jacobson's calculations can be found in \cite{18}. Jacobson used the Bekenstein-Hawking entropy and by substituting the Bekenstein-Hawking entropy into the Clausius equation,
\begin{equation}\label{eq12}
S_{BH} =( \frac{k_B c^3 A_0}{4\hbar G}), \hspace{.3cm} \ \partial s=\frac{\partial Q}{T}
\end{equation}
where $A_0$  is the area of the horizon, and by using the following relations
\begin{equation}\label{eq13}
\partial A_0 = - \int \lambda R_{ab} k^a k^b d\lambda dA
\end{equation}
\begin{equation}\label{eq14}
\partial Q = - \int \lambda T_{ab} k^a k^b d\lambda dA
\end{equation}
And some series of calculations  Einstein field equation is obtained as
\begin{equation}\label{eq15}
R_{ab} - \frac{1}{2} g_{ab} R + \Lambda g_{ab}  = \frac{8\pi G}{c^{4}}T_{ab}
\end{equation}
Now, if we use the entropy function in Eq \eqref {eq11} instead of Bekenstein-Hawking entropy into Clausius equation, we will get the following equation for gravity
\begin{equation}\label{eq16}
R_{ab} - \frac{1}{2} g_{ab} R + \Lambda g_{ab}  = \frac{4\pi G}{c^{4}}T_{ab}
\end{equation}
The only difference between Eq \eqref {eq15} and  Eq \eqref {eq16} are in r.h.s coefficients of the equations. Now if we expand the Eq \eqref {eq16} for weak fields, we get the following equation
\begin{equation}\label{eq17}
F = \frac{GmM}{2r^2}
\end{equation}
This equation is the same as with the equation that obtained in  \cite{29}, (in paper  \cite{29} the author derived  Equation \eqref {eq17} by  quantization of space and for classical limit $(\hbar \rightarrow0)$ ). And here again we got the same equation  by maximizing the entropy  equation in  Eq \eqref {eq6}. The coefficient of the entropy in Equation \eqref {eq11} and Newton's gravity equation  \eqref {eq17} shows that if we take twice value of $\langle A \rangle$ in Eq \eqref {eq4} instead of $\langle A \rangle$, the problem of the coefficient $(\frac{1}{2})$ in  Equation \eqref {eq11} and  Equation \eqref {eq17}(and  in paper \cite{29}) is solved
\begin{equation}\label{eq18}
A_0 =2 N\langle A \rangle 
\end{equation} 
Thus by using this relation, we get the following equations
\begin{equation}\label{eq19}
F = \frac{GmM}{r^2}
\end{equation}
\begin{equation}\label{eq20}
S_{max} = (\frac{k_B c^3 A_0}{2\hbar G}){ln(\frac{\sqrt{5}+1}{2})}  \simeq\ ( \frac{k_B c^3 A_0}{4\hbar G})
\end{equation}
Now, we whant to compute the error of two entropy equations, and we will show that the value of  maximum entropy  is very close to the Bekenstein-Hawking entropy,
\begin{equation}\label{eq21}
error=\frac{S_{BH} -S_{max}}{S_{BH}}= .037
\end{equation}
Where, we reach to a sharp and beautiful analogy that
\begin{equation}\label{eq21}
S_{max}\approx S_{BH}
\end{equation}
In summary, our calculations showed that the space quantas excitation can be the origin of gravity or in other words, gravity is nothing but the space quantas excitation. 
\section{Conclusion}
Studying the black hole thermodynamics and finding the origin of the black hole entropy is one of the most popular topics in the current era. The purpose of this paper is to present a very simple and efficient method for deriving The Bekenstein-Hawking entropy. And we showed that by maximizing the entropy of two-dimensional space, one can reach to the Bekenstein-Hawking entropy. On the other hand, our calculations showed that the excitation of space quanta at the microscopic scale causes gravity at the macroscopic scale, and so gravity is a quantum mechanical phenomenon. Summarizing, the above calculations showed that issues such as the origin of the black hole entropy and the cause of space-time bending have the same origin.
\bibliographystyle{ieeetr}
\bibliography{ref}

\end{document}